\documentclass{article}%
\usepackage{rjscmplx,graphics,epsf,apalike}
\usepackage{pstricks}
\usepackage{amsmath}
\usepackage{amsfonts}
\usepackage{amssymb}
\usepackage{graphicx}%
\title{Emergent Effective Collusion in an Economy of Perfectly Rational
  Competitors}
\author{Russell K. Standish\addrmark{1} and   Steve  Keen\addrmark{2}}

\address[1]{School of Mathematics, University of New South Wales, Australia}
\email[1]{r.standish@unsw.edu.au, http://parallel.hpc.unsw.edu.au/rks}

\address[2]{School of Economics and Finance, University of Western Sydney}
\email[2]{s.keen@uws.edu.au, http://www.debunkingeconomics.com}

\def\th{\mbox{$^\mathrm{th}$}}

\newlength\sblength
\def\slashbox#1#2{
  \psset{xunit=2.6em,yunit=3ex}
  \pspicture(-.75,-.75)(.75,1)
  \psline[linewidth=.5pt](-1,1)(1,-1)
  \rput(-.5,-.4){#1}
  \rput(.5,.4){#2}
  \endpspicture
}

\begin{document}

\maketitle

\begin{abstract}
We consider a simple model of rational agents competing in a single product
market described by simple linear demand curve. Contrary to accepted economic
theory, the agents' production levels synchronise in the absence of conscious
collusion, leading to a downward spiraling of market total production until
the monopoly price level is realised. This is in stark contrast to the
standard predictions of an ideal rational competitive market. Some form of
randomness in the form of agent irrationality, or non-synchronous updates is
needed to break this emergent \textquotedblleft collusion\textquotedblright%
\ and return the standard predictions of lower prices in a competitive market.

\end{abstract}

\section{Equilibrium Theory}

\label{equilibrium}

The standard economic theory of competitive markets argues that the behavior
of rational profit-maximizing firms leads to an equilibrium price that equals
the marginal cost of production.\cite{Mankiw04} The argument is put in the
following way.\ Firstly, profit for the \textit{i}th firm in an \textit{n}%
-firm industry is defined as:
\begin{equation}
\Pi(q_{i})=P(Q)q_{i}-TC(q_{i}), \label{ProfitDefinition}%
\end{equation}
where $\Pi(q_{i})$ is the profit of firm $i$ producing quantity $q_{i}$,
$P(Q)$ is the market price, which depends only on total production
($Q=\sum_{i}q_{i}$), also known as the \emph{demand curve}, and $TC(q_{i})$ is
the production cost incurred by firm $i$.

Assuming that $P^{\prime}(Q)<0$ (increasing supply reduces prices) and that
$TC^{\prime}(q)>0$, it is alleged that profit will be maximised when the
derivative of (\ref{ProfitDefinition}) is zero:
\begin{equation}
0=\Pi^{\prime}(q_{\pi_{\max}})=P(Q)+q_{\pi_{\max}}
  \left.\frac{\partial P}{\partial q_i}\right|_{q_i=q_{\pi_{\max}}}
  -TC^{\prime}(q_{\pi_{\max}}) \label{profit-derivative}%
\end{equation}

Economists describe $MR(q,Q)\equiv P(Q)+qP^{\prime}(Q)$ as
\textquotedblleft marginal revenue\textquotedblright\ and $MC(q)\equiv
TC^{\prime}(q)$ as \textquotedblleft marginal cost\textquotedblright.
In the standard Marshallian theory of markets, all firms are said to
maximize profits by equating marginal cost and marginal revenue,
independent of the number of firms in the industry---i.e., regardless
of whether the industry is a monopoly $(n=1)$, or \textquotedblleft
competitive\textquotedblright, by which it is meant that $n$ is so
large than the $i\th$ firm produces an infinitesimal proportion of
total industry output. Firms are also assumed to be managed
independently and not react strategically to each other's behavior, so
that $\partial q_{j}/\partial q_{i}=0,\forall i\neq j$ (this
assumption is known as \emph{atomism}).\footnote{This assumption is
  dispensed with in the game theoretic approach to competition, and
  replaced by the presumption of strategic responses by each firm to
  each other firm's output. However the non-strategic assumption is
  integral to Marshallian partial equilibrium analysis 
  where independent profit maximizing behavior is assumed. This
  Marshallian analysis remains the bedrock of instruction in
  economics, even though game theory now plays a substantial role in
  \textquotedblleft cutting edge\textquotedblright\ research.} A
difference is then alleged to exist between monopolies and competitive
industries. Marginal
revenue for the monopoly is:%
\begin{equation}
P(Q)+QP^{\prime}(Q)<P(Q) \label{MRMonopoly}%
\end{equation}
since $P^{\prime}(Q)<0$. Profit-maximizing behavior by a monopoly thus leads
to a market price that exceeds the marginal cost of production. On the other
hand, each firm in a competitive industry is said to act as a
\textquotedblleft price-taker\textquotedblright\ that cannot influence the
price in a market, so that $\partial P/\partial q_i =
P^{\prime}(Q)=0$. Marginal revenue for the $i$\th\ 
firm is thus:%
\begin{equation}
P(Q)+q_{i}P^{\prime}(Q)=P(Q) \label{MRCompetitive}.
\end{equation}
Eq (\ref{profit-derivative}) for the competitive firm therefore reduces to
\begin{equation}
P(Q)=TC^{\prime}(q_{\pi_\mathrm{max}}) \label{Cournot-individual},
\end{equation}
or in other words the equilibrium price equals marginal cost for the
$i$\th\ firm. Clearly, this can only be an equilibrium if all firms
face the same marginal costs, which implies that the aggregate
marginal cost of producing industry output $Q$ is equal to the
marginal cost faced by each firm producing $q_{i}$. Profit-maximizing by
individual firms leads to an aggregate output level at which price
equals marginal cost. We denote $Q_{C}$ as the total production level
in this case:\footnote{We use the subscript \textit{C} in $Q_{C}$ in
  reference to Cournot, who in 1838 first applied calculus to the
  issue of the
determination of output by profit-maximizing firms.}%
\begin{equation}
P(Q_{C})=TC^{\prime}(Q_{C}) \label{Cournot}%
\end{equation}

In contrast to this standard analysis, Stigler \citeyear(1957){Stigler57}
pointed out that $\partial P/\partial q_i$ is \textit{not} zero, but in fact equal to
$P^{\prime}(Q)$, ie $\frac{\partial P}{\partial q_{i}}=\frac{dP}{dQ}%
\frac{\partial Q}{\partial q_{i}}=\frac{dP}{dQ}$. Therefore
equation (\ref{Cournot-individual}) is incorrect and needs to be revised to:%
\begin{equation}
P(Q)-TC^{\prime}(q_{\pi_{\max}})=-q_{\pi_{\max}}P'(Q)>0.
\label{Cournot-corrected}%
\end{equation}
Despite its obvious mathematical correctness, most economics textbooks
(and economics researchers) do not use this result, but instead
continue to argue that $\partial P/\partial
q_{i}=0$.\footnote{Examples include advanced texts such as Varian
  \citeyear(1999){Varian99}, pp. 377--378, and Mas-Collell, {\em et
    al.} \citeyear(1995){Mas-Collell-etal95} pp. 315, 411--413 \& 661,
  in addition to almost all introductory economics texts (we know of
  no exceptions).}. This could be because, as well as pointing out the
incompatibility of the assumption of \textquotedblleft
price-taking\textquotedblright\ behavior ($\partial P/\partial
  q_{i}=0$) in the context of atomism ($\partial q_{j}/\partial
q_{i}=0,\forall i\neq j$), Stigler also proposed a reformulation of
marginal revenue in the symmetric case $q_{i}=q,\,\forall i$, that
appeared to describe eq (\ref{Cournot-individual}) as the limit of
profit-maximizing behavior as $n\rightarrow\infty$:
\begin{equation}
P(Q)+q_iP'(Q)=P\left(  1+\frac{1}{nE}\right)
\end{equation}
($E=\frac{P}{Q}\frac{dQ}{dP}$ is called the \emph{market elasticity of
demand}). Marginal revenue for the individual competitive firm therefore
converges to market price as $n\rightarrow\infty$. On the presumption that
profit maximizing behavior involves equating marginal revenue and marginal
cost, the $n$-firm economy approaches the Cournot equilibrium
(\ref{Cournot-individual}) as $n\rightarrow\infty$ (however, this also implies
that $q_{i}\rightarrow0$ in this same limit).

However, using Stigler's relation, Keen \citeyear(2003){Keen03a} shows that
equating marginal cost and marginal revenue \textit{does not} maximize profits
in a multi-firm industry. Keen's method was proof by contradiction: assuming
that all firms did equate marginal cost and marginal revenue, Keen showed that
this resulted in an aggregate output that exceeded the profit-maximizing
level. Therefore part of industry output had to be produced at a loss, which
in turn meant that some individual firms had to be producing beyond their
profit-maximizing output. Summarizing this analysis, if all firms equate
marginal revenue and marginal cost, then%
%
%
%
%
\begin{eqnarray}
\lefteqn{\sum_i\left(P(Q)+q_iP'(Q)-MC(q_i)\right) = 0} \nonumber\\
& \Rightarrow&  nP(Q) + QP'(Q)-nMC(Q) = 0 \nonumber\\
& \Rightarrow& MR(Q,Q)-MC(Q) = -(n-1)\left(P(Q)-MC(Q)\right),
\end{eqnarray}
which is negative by virtue of Stigler's relation (\ref{Cournot-corrected}).
This aggregate output level\ exceeds the profit maximizing level, since
aggregate marginal cost exceeds aggregate marginal revenue. This relation
quantifies the aggregation fallacy in the proposition that firms maximize
their profit by setting the \textit{partial} derivative of their profit with
respect to their own output to zero when in a multi-firm industry the
equilibrium profit level is found by setting the \textit{total} derivative of
each firm's profit to zero:%

%
%
%
%
%

\begin{equation}\label{total-derivative}
\frac{d\Pi(q_i)}{dQ} = \sum_j\frac{\partial\Pi(q_i)}{\partial
  q_j}\frac{\partial q_j}{\partial Q} = P(Q)+nq_iP'(Q)-MC(q_i) =
  0
\end{equation}
We can express this equation as 
\begin{equation}\label{q_iP'(Q)}
q_iP'(Q)=-\frac1n\left(P(Q)-MC(q_i)\right).
\end{equation}
and
\begin{equation}\label{marginal revenue}
P(Q)+q_iP'(Q)-MC(q_i) = -(n-1)q_iP'(Q)
\end{equation}
Substituting (\ref{q_iP'(Q)}) into (\ref{marginal revenue}), we arrive
at the following condition for the profit-maximizing equilibrium
(which we call the Keen equilibrium rather than the \textquotedblleft
monopoly\textquotedblright\ equilibrium as economists normally
describe it) \cite{Keen2004a,Keen2004b}. Denoting the total market
production in this case by $Q_K$, each firm's output satisfies:
\begin{equation}
P(Q_{K})+q_{i}P^{\prime}(Q_{K})-MC(q_{i})=\frac{n-1}{n}\left(
P(Q_{K})-MC(q_i)\right)  \label{Keen}%
\end{equation}

This formula corresponds to the accepted formula for a monopoly, but in the
$n$-firm case, firms produce at levels where their own-output marginal revenue
\textit{exceeds} their marginal cost. It is easily shown that this formula
maximizes profit for any total revenue and total cost functions that obey the
standard conditions $P^{\prime}(Q)<0$ and $TC^{\prime}(q)>0$.

In what follows, we use multi-agent modelling to (a) dispute the
\textquotedblleft\textit{given other firms' outputs fixed}\textquotedblright%
\ definition of rationality, (b) show that the Cournot equilibrium is
locally unstable while the Keen equilibrium is locally stable, (c)
establish that operationally rational profit-maximizing behavior leads
to the Keen equilibrium, and (d) quantify the degree of operationally
irrational behavior needed for the output to converge to the Cournot
equilibrium. The behavior of the agents and the overall system
illustrates an interesting phenomenon in multi-agent dynamics: the
emergence of apparently coordinated behavior even though agents are
not colluding or even communicating with each other.

\section{Dynamics}

With the assumption that $\partial P/\partial q_i=0$, eq
(\ref{profit-derivative}) asserts that each firm works in isolation.
Each firm considers what is the optimal level of production on the
assumption that the production levels of all other firms remain fixed.
Yet this is a contradictory situation, as every firm is attempting to
adjust its production levels to maximise its profit at the same time.
Rational firms, who are themselves adjusting their own outputs in a
search for the level that maximises profits, are in effect assumed to
irrationally believe that other firms are not doing the same thing.
Therefore in each time period, the total market production $Q$ cannot
be predicted.

While it is possible that aggregate historical data exists (production levels
and prices) that might allow a demand curve to be deduced, this is of little
help in the competitive setting. Rational agents performing optimisation
according to eq (\ref{profit-derivative}) cannot exist since each behaves as
if it has no impact on market price, and therefore no market price can
therefore be determined: the price would be what it always was ---
\textquotedblleft Que sera sera\textquotedblright\ --- without any tendency to
move. The actual equilibrium production values (if an equilibrium exists) must
arise in a dynamical setting where agents make decisions based on limited
knowledge and where, however small, their actions have an impact on their own
situation and therefore the aggregate market outcome. Obviously there is a
need to examine how real firms make decisions, however it is theoretically
instructive to consider the ideal case of simple firms making rational
decisions based on information available.

In this paper, we assume the only information a firm has is its own cost
function, the decision it made on the previous time step (increase or decrease
production), and what impact this decision had on its profit levels. The
obvious operationally rational use of this information is to continue changing
levels of production in the same direction if profits are rising, and to
reverse the direction of change if profits fall. To make the opposite
decision, to \textquotedblleft buck the trend\textquotedblright\ in other
words, is to discard any trend information that might be contained within the
profit time series.

It is difficult to follow analytically the outcome of more than two
interacting firms. The equilibrium theory described in \S\ref{equilibrium} is
assumed to be valid in the limit of an infinite number of firms. Therefore, we
resort to numerical simulations to explore what happens as the number of firms
is increased. In the following numerical simulations, we use affine demand and
cost rules:
\begin{align}
P(Q)  &  =a-bQ=100-10^{-2}Q\\
TC(q)  &  =cq=50q
\end{align}
For this model, the Cournot and Keen Equilibria can be computed from eq
(\ref{Cournot}) and (\ref{Keen}):
\begin{align}
Q_{C}=(a-c)/b=5000  &  ;P(Q_{C})=50\\
Q_{K}=(a-c)/2b=2500  &  ;P(Q_{K})=75
\end{align}
In this model, $Q_{K}$ corresponds to the monopoly production levels. The code
used for this simulation is available from the author's
website\footnote{Firmmodel.1.D3, from
http://parallel.hpc.unsw.edu.au/getaegisdist.cgi. It depends on a recent
version of
{\sffamily\slshape\mbox{\raisebox{.5ex}{Eco}\hspace{-.4em}{\makebox[.5em]{L}ab}}}%
{} being installed, which is also available from this website.}.

\begin{figure}
\begin{center}
\epsfbox{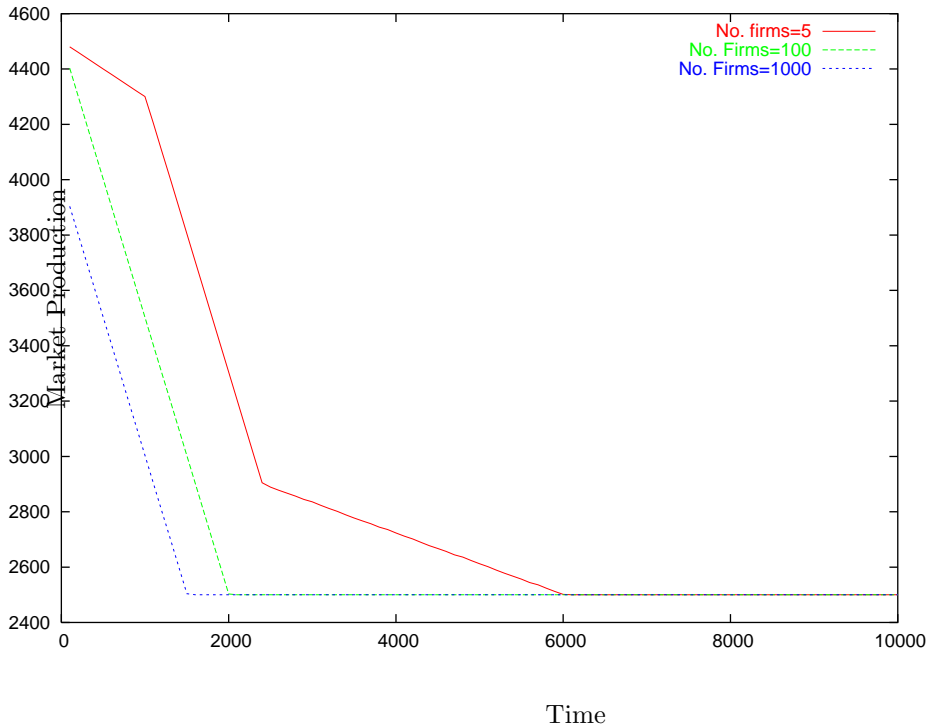}\mbox{}\\ Time
\end{center}
\par
\vspace{-7.5cm}%
\raisebox{4cm}{\rotatebox{90}{\rule{0pt}{8ex}Market Production}}\caption{Total
market production starting at $Q=4500$ for different sizes of economy. The
different slopes on the $n=5$ curve correspond initially to only 3 firms
synchronised, then 5 firms, then finally several firms overshoot their share
of the final production and begin to increase production, slowing the final
approach to equilibrium. The precise shape of the curve depends on the initial
distribution of increments, which is assigned randomly.}%
\label{nfirms}%
\end{figure}

Figure \ref{nfirms} shows the convergence to monopoly prices for a number of
different economy sizes. The simulations were started at the Cournot level in
each case, and rapidly converged to the monopoly levels.

It is instructive to consider the two firm case to understand what is going
on \cite{Simon-etal73}. Consider first the case where both firms have been
initialised to a production of $q>Q_{K}/2=(a-c)/4b$ with a negative increment
$\delta$. They will then both decrease production, leading to increased
profits for both firms.
\begin{align}
\Pi(q-\delta)  &  =(a-c)(q-\delta)-2b(q-\delta)^{2}\nonumber\\
&  =\Pi(q)-(a-c-4bq)\delta-2b\delta^{2}\\
&  >\Pi(q)
\end{align}
Since they are acting rationally, they will continue this action until global
profit maximisation is reached --- the monopoly situation.

If, however, both firms are initialised with a positive increment, both firms
will immediately increase production, leading to falling profit levels. Since
they are both acting perfectly rationally, they will then both switch to
negative increments, leading to increased profits which will converge on the
monopoly outcome as before.

Finally, consider the case of the firms being initialised with opposing
increments, with \samepage{$q < Q_{C}/2=(a-c)/2b$}:
\begin{align}
\Pi(q-\delta)  &  = (a-c)(q-\delta)-2bq(q-\delta)\nonumber\\
&  = \Pi(q) - (a-c-2bq)\delta\\
&  < \Pi(q)\nonumber\\
\Pi(q+\delta)  &  = (a-c)(q+\delta)-2bq(q+\delta)\nonumber\\
&  = \Pi(q) + (a-c-2bq)\delta\\
&  > \Pi(q).
\end{align}
Then the firm with the negative increment will switch to a positive increment
and we are back in the previous situation of both firms having the same
increment. Exactly the same situation pertains with $q>Q_{C}/2$, except that
the firm with positive increment switches to a negative increment. The end
result is monopoly levels.

Only in the case $q=Q_{C}/2=(a-c)/2b$ do production levels remain constant at
the Cournot level. It turns out that this ``metastable'' state is sensitive to
numerical rounding in the model. The probability of this happening in an
initialisation decreases as $2^{-n/2}$ as the number of firms $n$ increases,
and cannot happen for odd $n$.

The interesting thing to note is that it is the agent's rationality which
causes the two agents to lock into a spiral of lower production levels and
increased prices. If at any stage during this process, one agent were to ramp
up production, they could increase their own production at the expense of the
other party. The other party would quickly follow suit, and the two agents
would then reenter the downward spiral again. As a result, the interactive
topography of the profit landscape makes the Keen equilibrium locally stable
for the two agents, while the Cournot equilibrium is locally unstable. This
can be seen by considering the impact of changes in output on profit levels
for the two firms in the vicinity of these equilibria. Tables
\ref{KeenProfitChange} and \ref{CournotProfitChange}
show the impact of changing output by one and two units by each firm in the
vicinity of the Keen and Cournot equilibria respectively. The only changes
which are viable are ones in which both profit change entries are positive,
since where a negative change occurs the firm experiencing this change will
reverse its direction of output change. As can be seen from Table
\ref{KeenProfitChange}, there is no change in output combination which is
viable: a change in any direction by either firm results in at least one
change in profit figure that is negative.\ Therefore the firm experiencing
this change will alter its direction of output change, thus returning the
collective output level to the Keen equilibrium figure.%

\begin{table}
\begin{center}
\begin{tabular}{|c|r|c|c|c|c|c|}
\hline
&\multicolumn1{|c}{}& \multicolumn5{c|}{Firm B}\\\hline
&\slashbox{$\Delta q_A$}{$\Delta q_B$}& $-2$ & $-1$ & 0 & 1 & 2 \\\cline{2-7}
& $-2$ & \slashbox{$-0.08$}{$-0.08$} & \slashbox{$-13$}{$-12$} &
\slashbox{$-25$}{$-25$} & \slashbox{$-38$}{38} & \slashbox{$-50$}{50}
\\\cline{2-7}
\rotatebox{90}{\makebox[0pt]{Firm A}}& $-1$ & \slashbox{$12$}{$-13$} & \slashbox{$-0.02$}{$-0.02$} &
\slashbox{$-13$}{$13$} & \slashbox{$-25$}{25} & \slashbox{$-37$}{37}
\\\cline{2-7}
& $0$ & \slashbox{$25$}{$-25$} & \slashbox{$13$}{$-13$} &
\slashbox{$0$}{$0$} & \slashbox{$-13$}{12} & \slashbox{$-25$}{25}
\\\cline{2-7}
& $1$ & \slashbox{$38$}{$-38$} & \slashbox{$25$}{$-25$} &
\slashbox{$12$}{$-13$} & \slashbox{$-0.02$}{$-0.02$} & \slashbox{$-13$}{12}
\\\cline{2-7}
& $2$ & \slashbox{$50$}{$-50$} & \slashbox{$37$}{$-37$} &
\slashbox{$25$}{$-25$} & \slashbox{$12$}{-13} & \slashbox{$-0.08$}{$-0.08$}
\\\hline
\end{tabular}
\end{center}
\caption{Table showing the change in profits for two companies around
  the Keen equilibrium, for $\Delta q_i=\pm1,\pm2$.}
\label{KeenProfitChange}
\end{table}

On the other hand, there is a viable direction of output change from
the Cournot equilibrium shown in Table \ref{CournotProfitChange}: if
both firms reduce their output, then both will experience an increase
in profits. Hence a reduction in output by both firms is a viable
strategy for each firm acting independently, and the firms will
therefore move away from the\ Cournot equilibrium. The Cournot
equilibrium is
therefore locally unstable, while the Keen equilibrium is locally stable.%

\begin{table}
\begin{center}
\begin{tabular}{|c|r|c|c|c|c|c|}
\hline
&\multicolumn1{|c}{}& \multicolumn5{c|}{Firm B}\\\hline
&\slashbox{$\Delta q_A$}{$\Delta q_B$}& $-2$ & $-1$ & 0 & 1 & 2 \\\cline{2-7}
& $-2$ & \slashbox{$33$}{$33$} & \slashbox{$17$}{$33$} &
\slashbox{$-0.04$}{$33$} & \slashbox{$-17$}{33} & \slashbox{$-33$}{33}
\\\cline{2-7}
\rotatebox{90}{\makebox[0pt]{Firm A}}& $-1$ & \slashbox{$33$}{$17$} & \slashbox{$17$}{$17$} &
\slashbox{$-0.01$}{$17$} & \slashbox{$-17$}{17} & \slashbox{$-33$}{17}
\\\cline{2-7}
& $0$ & \slashbox{$33$}{$-0.04$} & \slashbox{$17$}{$-0.01$} &
\slashbox{$0$}{$0$} & \slashbox{$-17$}{-0.01} & \slashbox{$-33$}{$-0.04$}
\\\cline{2-7}
& $1$ & \slashbox{$33$}{$-17$} & \slashbox{$17$}{$-17$} &
\slashbox{$-0.01$}{$-17$} & \slashbox{$-17$}{$-17$} & \slashbox{$-33$}{$-17$}
\\\cline{2-7}
& $2$ & \slashbox{$33$}{$-33$} & \slashbox{$17$}{$-33$} &
\slashbox{$-0.04$}{$-33$} & \slashbox{$-17$}{$-33$} & \slashbox{$-33$}{$-33$}
\\\hline
\end{tabular}
\end{center}
\caption{Table showing the change in profits for two companies around
  the Cournot equilibrium, for $\Delta q_i=\pm1,\pm2$.}
\label{CournotProfitChange}%
\end{table}

The $n$-firm case is similar. It only takes more than 50\% of the agents to be
simultaneously decreasing production to reduce the total market production,
which will lead to increased profits for all. Even though none of the agents
are communicating with each other, the effect is as if they were colluding
with each other to keep market prices high.

\section{Variant Dynamics}

\subsection{Irrational Dynamics}

\begin{figure}
\centerline{\epsfbox{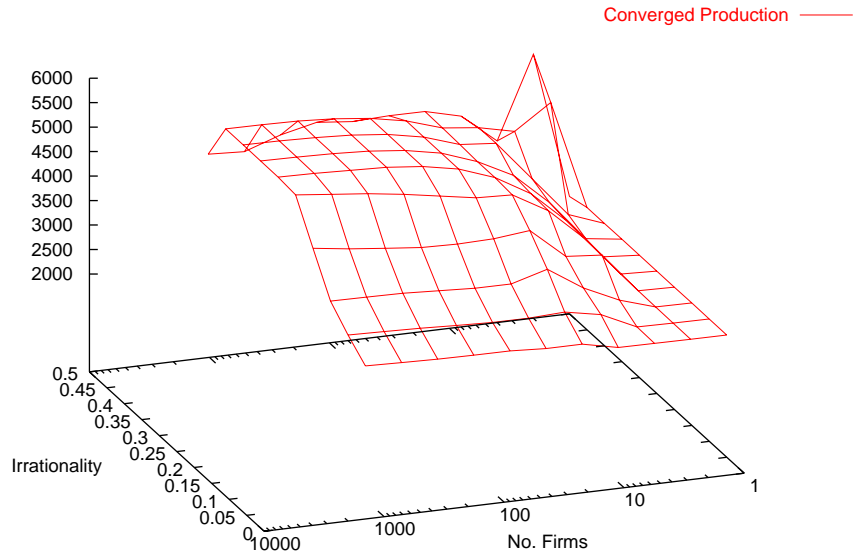}}\caption{Plot of the converged production
levels as a function of irrationality and number of firms. Converged
production level is computed as the average production value $Q$ between
$t=40000$ and $t=50000$.}%
\label{irrat2D}%
\end{figure}

Figure \ref{irrat2D} plots the effect of irrationality versus number of firms.
Irrationality governs the probability that an agent makes the opposite
decision to the rational one. It can be seen that quite high levels of
irrationality are required to break the trend towards monopoly valued
economies. There is a fairly sharp transition between the system being pulled
to the Keen equilibrium in the rational case, and the Cournot equilibrium
when irrationality lies between 0.3 and 0.5 for many firms. If the firms are
more than 50\% irrational then neither equilibrium is an attractor, as firms
are no longer profit seeking.

\subsection{Update rules}

\begin{figure}
\centerline{\epsfbox{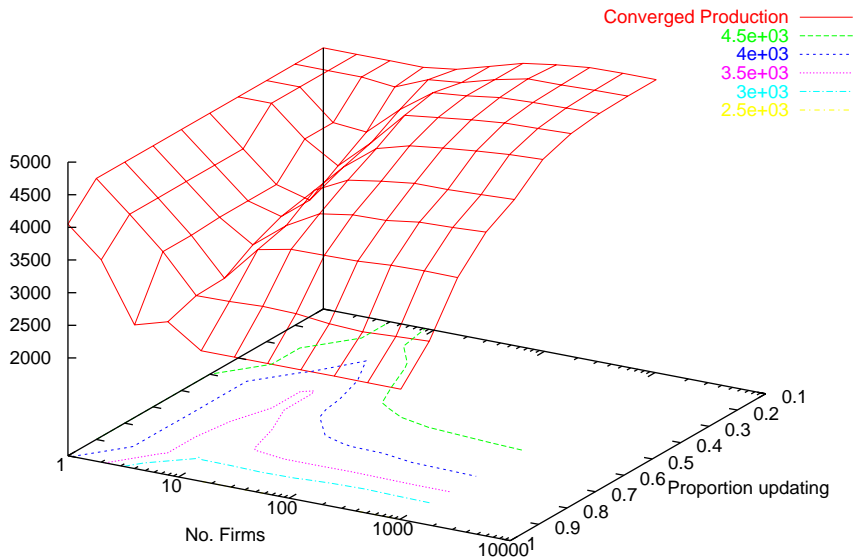}}\caption{Plot of the converged production
levels as a function of the number of firms, and the proportion of those firms
updating their production levels each timestep. Converged production level is
computed as the average production value $Q$ between $t=40000$ and $t=50000$.}%
\label{update}%
\end{figure}

Until now we've been considering the effect of all firms updating production
levels every timestep. The opposite extreme of only one firm updating its
production level leads to the ideal situation considered for the Cournot
equilibrium. Interpolating between the two cases, we can consider dynamics
where only a fraction of firms update there production rules in any timestep,
selected at random from the list of all firms. Figure \ref{update} shows the
converged production values as a function of the proportion of firms updating
each timestep, and number of firms.

This outcome occurs because when each firm alters its output while other firms
hold theirs constant, the change in revenue to which the firm reacts is its
own-output marginal revenue, as defined above. This will indicate to the firm
that it can increase its profit by increasing its output from the Keen level,
in a process which terminates at the\ Cournot equilibrium.

However, this behavior is both unrealistic (other firms do not hold their
output while one firm varies its in the real world), and involves a peculiar
form of irrationality---since though the change in output is done with the
intention of increasing profit, over each iteration it results in the firm's
profit level falling, once all other firms have done likewise. Thus what
appears rational within each iteration becomes irrational from one iteration
to the next. A less myopic behavioral rule---which included consideration of
the profit outcome from one iteration to the next---would probably reverse
this behavior. In this work, only the result from the last and last but one
market outcome is used in determining the agents' behaviour. It is, of course,
clearly possible to make use of more market historical data, say the results
of the last $\ell>2$ market periods. It is not clear, in this case, what the
response of a rational agent should be. One possibility is a lookup decision
table indexed by the relative movements of the market over the last $\ell$
periods, analogous to the \emph{minority game} \cite{Challet-Zhang97}.

The stability of this myopic sequential update rule is also dependent
on the percentage of firms that change output at each timestep. For
large economies ($>32$ firms), the system converges to the Cournot
equilibrium (5000) when only 10\% of the firms are updating their
production levels per timestep, and to the Keen equilibrium when all
firms are updating their production levels. The final production
levels shows a smooth decline as the percentage of firms updating
increases.

The somewhat strange behaviour at small economy sizes can be explained by the
fact that individual firms have proportionately larger effects on total
production values, so only a few firms need to synchronise to pull production
levels down to monopoly values. In the single firm case, no firms are updating
at all, so final values are simply the initial values of 4500.

\section{Conclusion}

The simulation results demonstrate that markets can be locked in a spiral of
restricted production converging on monopoly pricing levels as though the
firms were colluding, \emph{even though} no interaction between firms takes
place. Perfect profit-seeking rationality and the dynamics of the economy
tends to lock firms into synchronous behaviour, leading to a global monopoly
behaviour. Only by inserting some randomness in the system by way of
irrationality, or by decoupling the firm's production periods, does the
economy ``melt'' into the behaviour predicted by competition theory. Clearly
such factors may operate in real markets, and this work does not suggest that
real competitive markets will produce monopolistic outcomes, but it does
suggest that the ideal perfectly rational competitive market does not produce
the expected competitive outcomes.

\section{Acknowledgments}

The authors would like to thank the \emph{Australian Centre for Advanced
Computing and Communications} for access to a large Linux cluster that enabled
the simulations to be completed in a timely manner.

\bibliographystyle{apalike}
\bibliography{rus}

\end{document}